\documentclass{article}%
\usepackage{amsfonts}
\usepackage{amsmath}
\usepackage{graphicx, color}%
\usepackage{amssymb}%
\usepackage{graphicx}
\providecommand{\U}[1]{\protect\rule{.1in}{.1in}}

\begin{document}

\title{A mathematical model for gene evolution after whole genome duplication}
\author{Y. Nakamura\thanks{yojnakam@affrc.go.jp}\\National Research Institute of Fisheries Science\\Japan Fisheries Research and Education Agency\\Yokohama, Kanagawa, Japan}
\maketitle

\begin{abstract}
Whole genome duplication (WGD) is one of the most important events in the
molecular evolution of organisms. In fish species, a WGD is considered to have
occurred in the ancestral lineage of teleosts. Recent comprehensive ortholog
comparisons among teleost genomes have provided useful data and insights into
the fate of redundant genes generated by WGD. Based on these data, a
mathematical model is proposed to explain the evolutionary scenario of genes
after WGD. The model is parameterized taking into account an equilibrium
between i) rapid loss of either of the duplicate genes and ii) moderate
functional differentiation of each of duplicate genes, both of which are
followed by slow gene loss under purifying selection. This model predicts
that, in the teleost lineage, a maximum of about 3000 gene pairs may have
differentiated functionally during 90 million years after WGD. Thus, the
present study provides a possibility that the whole impact of WGD can be
quantitatively assessed according to the model parameters, before details of
genomic structural changes or functional differentiation are investigated. If
the equilibrium model is valid not only for teleosts but also for other
lineages that have undergone WGDs, correlations between the assessment indices
and evolutionarily significant events, such as the diversification of species
or the occurrence of novel phenotypes, could be tested and compared among
those lineages.

\end{abstract}

\section{Introduction}

In the field of genetics, much attention has been given to gene duplication
and its significance in the evolution of organisms \cite{ref1,ref2,ref3}.
Whole genome duplication (WGD), in view of its large scale, is considered to
be one of the most significant evolutionary events. A well-known example is
that the ancestor of vertebrate species went through at least two WGDs
\cite{ref2,ref4} 500--800 million years ago \cite{ref5}. In the lineage of
fish, one more WGD (teleost-specific WGD or TS-WGD) is estimated to have
occurred in the ancestor of teleosts 300--400 million years ago
\cite{ref5,ref6,ref7}. WGD generates functionally identical copies of genes in
a genome, resulting in a situation in which either of the duplicate genes
becomes a spare or dispensable gene. It is considered that, after such an
expansion of genes by WGD, many of the redundant partners become pseudogenes
by degenerative mutations \cite{ref3,ref8} and will be finally lost in the
course of evolution. Simultaneously, for some gene pairs, the functional
redundancy may be lost by chance due to some mutations. In such cases, either
or both of the duplicate genes become to have novel roles that are different
from the original one \cite{ref2,ref8}, and thereby both of them become to
evolve mainly under purifying selection. Actually, many cases of the
functional differentiation in duplicate genes have been reported \cite{ref1}.
In addition, many theoretical studies have investigated gene duplication from
the view of population genetics \cite{ref9,ref10,ref11,ref12}, but studies
based on the assessment of large-scale data have not been fully done.
Recently, a quantitative comparison among teleost genes was performed at the
genomic level, which provided significant implications about the evolutionary
fate of duplicate genes after WGD \cite{ref13}. In particular, the result
showed that the loss of redundant genes generated by TS-WGD was very rapid
with more than 70\% of gene pairs becoming single during 60 million years,
after which the rate of gene loss was very slow. In the present study, I
propose a simple mathematical model to explain the gene loss patterns observed
after WGD. The model is parameterized taking into account an equilibrium
between two evolutionary scenarios for duplicate genes: i) rapid loss under
functional dispensability to each other, and ii) moderate occurrence of
functional differentiation to each other. Additionally, slow gene loss in a
conservative manner, in which purifying selection is dominant, is
parameterized in the model. In this study, I applied the equilibrium model to
the recent data derived from a comparison of teleost genes at the genomic
level \cite{ref13}, showed that the model explained the data well, and
discussed the potential of the model to assess the impact of WGD on all the
genes in the organisms which went through WGDs.

\section{Mathematical model}

In the equilibrium model, three parameters are defined: $\alpha$, for the rate
of loss of a functionally redundant partner in a duplicate gene pair; $\beta$,
for the rate of functional differentiation in a duplicate gene pair; and
$\gamma$, for the rate of loss of non-redundant genes. Two types of
non-redundant genes are considered: i) single genes that have lost their
partners and ii) genes that have functionally differentiated in a pair. Thus,
$\gamma$ is associated with the normal process of gene evolution under
purifying selection, and the value can also be computed in other studies
regardless of the context of gene duplication. It should be noted that
\textquotedblleft functional differentiation\textquotedblright\ events can
occur through mechanisms such as neofunctionalization, subfunctionalization,
or dosage selection \cite{ref14}. Such mechanisms are not distinguished in the
model; all of them may be included in the value of $\beta$, and
\textquotedblleft functional differentiation\textquotedblright\ is rather
defined as any event in which the type of natural selection acting on genes is
switched (i.e., from relaxed selection to purifying selection). In the model,
I assumed that the loss of duplicate genes occurs one-by-one; that is, both
duplicate genes are not lost at the same time. All the possible states of a
gene pair and the corresponding model parameters are summarized in Figure 1.%
\begin{figure}
[ptb]
\begin{center}
\includegraphics[
height=2.2131in,
width=3.7446in
]%
{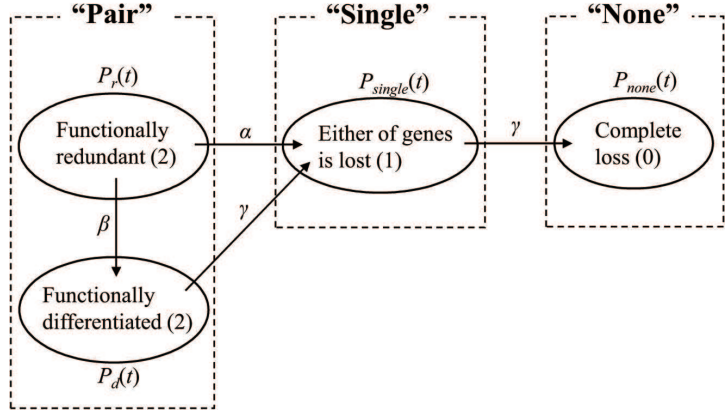}%
\caption{{\protect\small All possible states of a duplicate gene pair in the
equilibrium model. The state of \textquotedblleft pair\textquotedblright\ is
composed of two states, \textquotedblleft functionally
redundant\textquotedblright\ and \textquotedblleft functionally
differentiated.\textquotedblright\ Each gene pair starts from the first state
\textquotedblleft functionally redundant\textquotedblright\ immediately after
WGD, and then transitions to other states. The numbers in parenthesis indicate
the numbers of genes remaining in a pair (0, 1, or 2). The parameters,
}$\alpha${\protect\small , }$\beta${\protect\small , and }$\gamma
${\protect\small \ indicate the rates of transition from one state to another,
and each of }$P_{r}(t)${\protect\small , }$P_{d}(t)${\protect\small ,
}$P_{single}(t)${\protect\small , and }$P_{none}(t)$%
{\protect\small \ indicates the probability that a gene pair is in the
corresponding state at time }$t${\protect\small .}}%
\end{center}
\end{figure}
In addition, I defined three probabilities, $P_{pair}(t)$, $P_{single}(t)$,
and $P_{none}(t)$, where $P_{pair}(t)$ is the probability that both duplicate
genes in a pair remain at time $t$ (state = \textquotedblleft
pair\textquotedblright), $P_{single}(t)$ is the probability that either of the
duplicate genes in a pair is lost at time $t$ (state = \textquotedblleft
single\textquotedblright), and $P_{none}(t)$ is the probability that both
duplicate genes in a pair are lost at time $t$ (state = \textquotedblleft
none\textquotedblright). Here, $t=0$ is the time point at which WGD occurred.
In the state of \textquotedblleft pair,\textquotedblright\ each of the gene
pairs is in either of two states, \textquotedblleft functionally
redundant\textquotedblright\ and \textquotedblleft functionally
differentiated\textquotedblright\ (Figure 1), and hence $P_{pair}(t)$ is given
by%
\[
P_{pair}(t)=P_{r}(t)+P_{d}(t).
\]
\linebreak In addition, these probabilities satisfy the following differential
equations:%
\begin{align*}
\frac{dP_{r}(t)}{dt}  & =-\left(  \alpha+\beta\right)  P_{r}(t),\\
\frac{dP_{d}(t)}{dt}  & =\beta P_{r}(t)-\gamma P_{d}(t),\\
\frac{dP_{single}(t)}{dt}  & =\alpha P_{r}(t)+\gamma\left(  P_{d}%
(t)-P_{single}(t)\right)  ,\\
\frac{dP_{none}(t)}{dt}  & =\gamma P_{single}(t),\\
& \left(  0<\alpha,\beta,\gamma<1,\alpha+\beta<1\right)
\end{align*}
\linebreak where $P_{r}(0)=1$ and $P_{d}(0)=P_{single}(0)=P_{none}(0)=0$.
Solving these equations, $P_{pair}(t)$, $P_{single}(t)$ and $P_{none}(t)$ are
represented as follows:%
\begin{align*}
P_{pair}(t)  & =\frac{\left(  \alpha-\gamma\right)  e^{-\left(  \alpha
+\beta\right)  t}+\beta e^{-\gamma t}}{\alpha+\beta-\gamma},\\
P_{single}(t)  & =\frac{1}{\left(  \alpha+\beta-\gamma\right)  ^{2}}[\{\left(
\alpha+\beta\right)  \left(  \alpha-\gamma\right)  +\beta\gamma\left(
\alpha+\beta-\gamma\right)  t\}e^{-\gamma t}\\
& -\left(  \alpha+\beta\right)  \left(  \alpha-\gamma\right)  e^{-\left(
\alpha+\beta\right)  t}],\\
P_{none}(t)  & =1+\frac{1}{\left(  \alpha+\beta-\gamma\right)  ^{2}}%
[\gamma\left(  \alpha-\gamma\right)  e^{-\left(  \alpha+\beta\right)  t}\\
& -\{\left(  \alpha+\beta\right)  ^{2}-\gamma\left(  \alpha+2\beta\right)
+\beta\gamma\left(  \alpha+\beta-\gamma\right)  t\}e^{-\gamma t}].
\end{align*}
\linebreak It should be noted that $P_{pair}$ converges to $\beta/\left(
\alpha+\beta\right)  $ at $t=\infty$ when $\gamma=0$, indicating that the loss
of gene pairs and the functional differentiation of duplicate genes have
reached an equilibrium state. In this study, however, $\gamma$ is larger than
zero; therefore, $P_{pair}$ continues to decrease and finally converges to
zero at $t=\infty$.

\section{Results and discussion}

The equilibrium model was applied to the teleost fish data published by Inoue
et al. \cite{ref13}. These data include information for duplicate genes in a
total of 6892 pairs, which were chosen based on a comparison among teleost and
outgroup genome data \cite{ref15}. The genes in these pairs are orthologous
among nine teleosts (Mexican tetra, zebrafish, Atlantic cod, Nile tilapia,
platyfish, medaka, stickleback, greenpuffer, and fugu), and the conservation
or loss of duplicate genes in each of the genomes is recorded in the original
data (Figure 2A).%
\begin{figure}
[ptb]
\begin{center}
\includegraphics[
height=4.0966in,
width=4.1347in
]%
{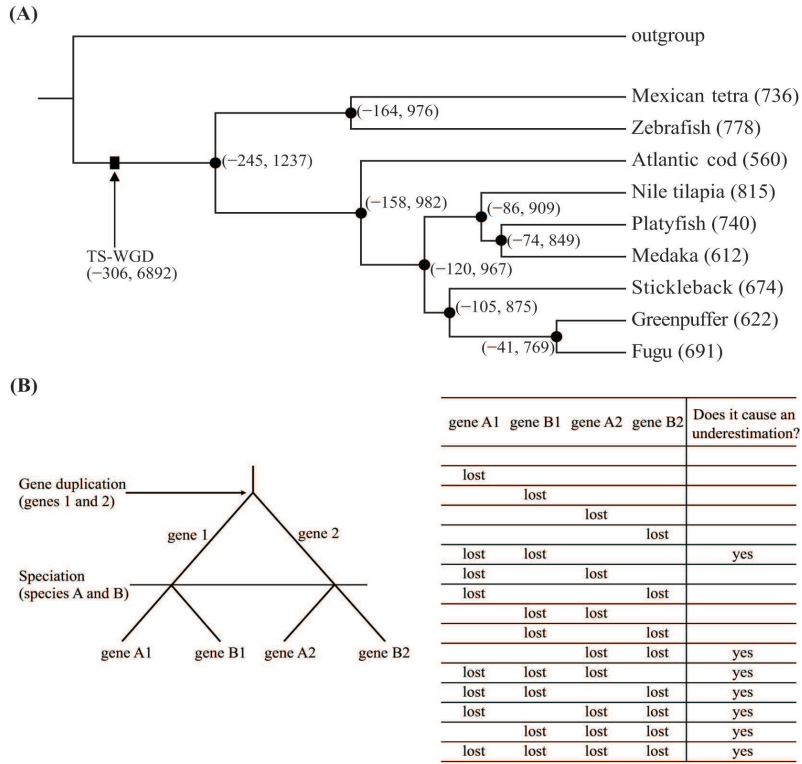}%
\caption{{\protect\small Phylogenetic relationship among the nine teleosts
examined in this study. (A) Phylogenetic tree of the nine teleosts examined.
At each node, the divergence time (million years) and the estimated number of
gene pairs in the state \textquotedblleft pair\textquotedblright\ are shown in
parenthesis. The time of the TS-WGD and the number of gene pairs are also
shown. The numbers next to the teleost names indicate the numbers of gene
pairs in the state \textquotedblleft pair\textquotedblright\ at present
(}${\protect\small t=0}${\protect\small ); the average is 692. All the data
are according to Inoue et al. \cite{ref13}. (B) Phylogenetic relationship of
duplicate genes (genes 1 and 2) followed by the divergence of species A and B
(left), and the patterns of gene loss causing an underestimation of gene pairs
in the state \textquotedblleft pair\textquotedblright\ (right).}}%
\end{center}
\end{figure}
The parameters $\alpha$, $\beta$, and $\gamma$ were fitted using the equation
of $P_{pair}(t+306)$ according to the numbers of gene pairs that were
estimated to have been present or were now present at 10 time points (i.e.,
nodes or edges in the phylogenetic tree)
($t=-306,-245,-164,-158,-120,-105,-86,-74,-41$, and $0$ million years,
$t=-306$ is the time of the TS-WGD, and $t=0$ is the present). From the
original data, I counted the numbers of gene pairs that were completely lost
in each of the nine teleost genomes; the average was $1191\pm311$ pairs (Table
1). The fitting was done by the primal-dual interior point method \cite{ref16}
implemented in Mathematica ver. 11 (Wolfram Research, Illinois, USA) under the
constraint of $P_{none}(306)=1191/6892$. As a result, $\alpha$, $\beta$, and
$\gamma$ were estimated to be 0.044, 0.0076, and 0.00078, respectively. The
behavior of $P_{pair}$ was well matched to that of the actual data (Figure 3A)%
\begin{figure}
[ptb]
\begin{center}
\includegraphics[
height=3.1427in,
width=4.1693in
]%
{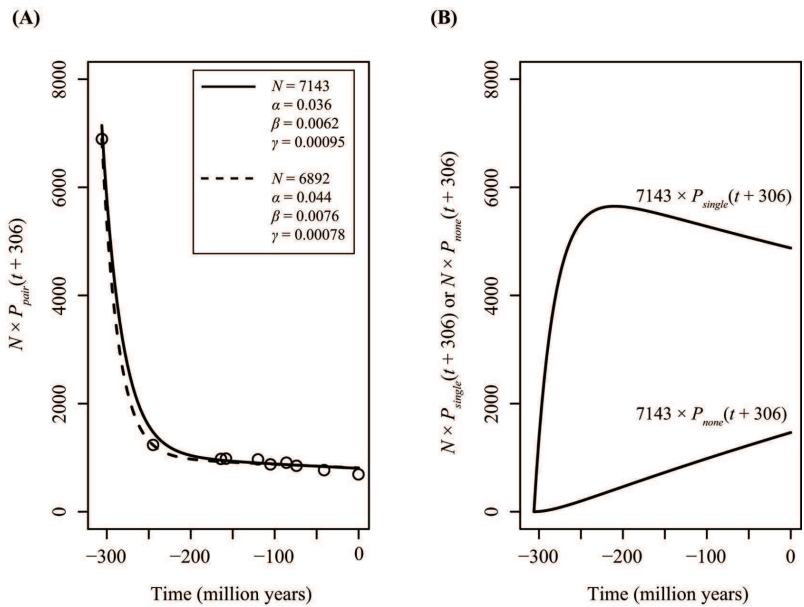}%
\caption{{\protect\small Estimated numbers of gene pairs after TS-WGD. (A)
Estimated numbers of gene pairs in the state \textquotedblleft
pair\textquotedblright\ (}${\protect\small N\times P}_{{\protect\small pair}%
}{\protect\small (t+306)} ${\protect\small ) with uncorrected (dashed line)
and corrected (solid line) parameters. The data shown in Figure 2A are also
plotted (open circles). (B) Estimated numbers of gene pairs in the state
\textquotedblleft single\textquotedblright\ (}${\protect\small N\times
P}_{{\protect\small single}}{\protect\small (t+306)}${\protect\small ) or
\textquotedblleft none\textquotedblright\ (}${\protect\small N\times
P}_{{\protect\small none}}{\protect\small (t+306)}${\protect\small ) with
corrected parameters.}}%
\end{center}
\end{figure}
and comparable to that of a recent model \cite{ref13}, suggesting that the
equilibrium model is a worthy alternative model. In particular, the
equilibrium model is consistent with the observation that a substantial number
of gene pairs ($1191/6892=17\%$ in average) were lost in the extant teleost
genomes. In the previous model, single genes that had lost their partners were
assumed to be indispensable for the teleost; therefore, single genes will
never be further lost. Such an assumption seems to be inconsistent with the
actual data.\newline%
\begin{tabular}
[c]{ll}%
\multicolumn{2}{l}{{\small Table 1. Number of gene pairs lost in extant
teleosts.}}\\\hline
{\small Teleost examined} & {\small Number of lost gene pairs}\\\hline
{\small Mexican tetra} & ${\small 938}$\\
{\small Zebrafish} & ${\small 739}$\\
{\small Atlantic cod} & ${\small 1498}$\\
{\small Nile tilapia} & ${\small 780}$\\
{\small Platyfish} & ${\small 1040}$\\
{\small Medaka} & ${\small 1423}$\\
{\small Stickleback} & ${\small 1237}$\\
{\small Greenpuffer} & ${\small 1653}$\\
{\small Fugu} & ${\small 1407}$\\\hline
{\small Average}$\pm${\small SD} & ${\small 1191\pm311}$\\\hline
\multicolumn{2}{l}{{\footnotesize SD, standard deviation.}}%
\end{tabular}
\newline

The estimated $\alpha$, $\beta$, and $\gamma$ parameters were further
corrected taking into account a feature of the original data, namely the data
were composed of genes that are still present in at least one of the nine
teleost genomes. For example, gene pairs that were completely lost during the
period of $-306<t<-245$ are never transmitted to the teleost genomes examined
(Figure 2A); therefore, these pairs should not be counted in the original
data. In addition, parallel gene losses in descending sister lineages after
the divergence at $t=-245$ will make the state \textquotedblleft
pair\textquotedblright\ untraceable, resulting in the underestimation of gene
pairs (Figure 2B). First, the true number of gene pairs to be observed at the
time of TS-WGD was defined as $N$. For the extant gene pairs in the original
data, the equation is%
\[
N\left(  P_{pair}(306)\right)  \simeq692.
\]
Next, two conditional probabilities, $D_{1}(t_{1},t_{2})$ and $D_{2}%
(t_{1},t_{2})$, were defined: i) the probability that when a gene pair is in
the state \textquotedblleft pair\textquotedblright\ at time $t_{1}$, either of
the duplicate genes will be lost at time $t_{2}$; and ii) the probability that
when a gene pair is in the state \textquotedblleft pair\textquotedblright\ at
time $t_{1}$, both duplicate genes in the pair will be lost at time $t_{2}$.
These conditional probabilities are given by%
\begin{align*}
D_{1}(t_{1},t_{2})  & =\frac{P_{r}(t_{1})\left(  1-e^{-\alpha\left(
t_{2}-t_{1}\right)  }\right)  +P_{d}(t_{1})\left(  1-e^{-\gamma\left(
t_{2}-t_{1}\right)  }\right)  }{P_{pair}(t_{1})},\\
D_{2}(t_{1},t_{2})  & =\frac{P_{none}(t_{2})-P_{none}(t_{1})-P_{single}%
(t_{1})\left(  1-e^{-\gamma\left(  t_{2}-t_{1}\right)  }\right)  }%
{P_{pair}(t_{1})},
\end{align*}
\linebreak where $t_{1}<t_{2}$. Using these probabilities, the ratio of gene
pairs that will be underestimated by parallel gene losses between a node of
time $t$ and two descending sister nodes or edges $a$ and $b$ (times, $t_{a}$
and $t_{b}$) is given by%
\begin{align*}
Q\left(  t,t_{a},t_{b}\right)   & =D_{2}(t,t_{a})D_{2}(t,t_{b})+D_{1}%
(t,t_{a})D_{2}(t,t_{b})\\
& +D_{2}(t,t_{a})D_{1}(t,t_{b})+\frac{D_{1}(t,t_{a})D_{1}(t,t_{b})}{2},
\end{align*}
where $t<t_{a}$ and $t<t_{b}$. Note that there are seven patterns of parallel
gene loss causing the underestimation of gene pairs in the state
\textquotedblleft pair\textquotedblright\ (Figure 2B), which correspond to one
of $D_{2}(t,t_{a})D_{2}(t,t_{b})$, two of $D_{1}(t,t_{a})D_{2}(t,t_{b})/2 $,
two of $D_{2}(t,t_{a})D_{1}(t,t_{b})/2$, and two of $D_{1}(t,t_{a}%
)D_{1}(t,t_{b})/4$. In the case of the last nodes ($t=-164,-74$, and $-41$),
which are followed by edges, the ratio of underestimation by parallel gene
loss is equal to $Q(t+306,306,306)$. Therefore, the numbers of gene pairs
counted in these nodes were corrected to $N\{1-Q(t+306,306,306)\}P_{pair}%
(t+306)$. Contrastingly, in the case of the deeper nodes
($t=-245,-158,-120,-105,$ and $-86$), which are followed by at least one node,
the patterns of parallel gene loss are much more complicated. For the deeper
nodes, I first performed the simulation of gene loss with uncorrected
parameters, then computed the ratio of underestimation by parallel gene losses
(Table 2). The results showed that, as in the case of the last nodes, the
ratio of gene pairs that will be underestimated by parallel gene losses could
be roughly approximated by $Q(t+306,t_{a}+306,t_{b}+306)$. Note that the
approximation may depend on the number of species examined or the phylogenetic
relationship. When other data sets are used, the method of parameter
correction may have to be modified. In the present study, the ratio of
underestimation at the deepest node ($t=-245$) was different by 5\% from
$Q(61,142,148)$, probably because the original data were sparse around this
node (Figure 3A). Letting the ratio of underestimation of gene pairs at
$t=-245$ be $Q_{-245}$, the formula $Q(61,142,148)<Q_{-245}<Q(61,306,306)$ is
apparently established, where $Q(61,306,306)$ is the ratio of underestimation
focusing on only two distantly related species (e.g., zebrafish and medaka).
Therefore, $Q_{-245}$ was approximated by $\{Q(61,142,148)+Q(61,306,306)\}/2$
in this study. Finally, for the period $-306<t<-245$, the losses of gene pairs
until the next two nodes ($t=-164$ and $-158$,) were taken into account, and
the proportion of gene pairs to be unobservable in the original data was
approximated by%
\begin{align*}
R  & =P_{none}(61)+P_{pair}(61)D_{2}(61,142)D_{2}(61,148)\\
& +P_{single}(61)\left(  1-e^{-\gamma\left(  142-61\right)  }\right)  \left(
1-e^{-\gamma\left(  148-61\right)  }\right)  .
\end{align*}
\linebreak The value of $R$ was about 0.031 with the uncorrected parameters,
close to the simulated estimate (Table 2). Thus, the following equations were
obtained:%
\begin{align*}
N\left(  1-R\right)   & \simeq6892,\\
N\left(  P_{none}(306)-R\right)   & \simeq1191.
\end{align*}
According to these equations, the four parameters were fitted again by the
Newton-Raphson method with initial values of $N$ = 6892, $\alpha$ = 0.044,
$\beta$ = 0.0076, and $\gamma$ = 0.00078. As a result, $N$ was estimated to be
7143, and $\alpha$, $\beta$, and $\gamma$ were corrected to 0.036, 0.0062, and
0.00095, respectively. The values of $\alpha$, $\beta$, and $\gamma$ increased
or decreased by about 20\%, and the total number of gene pairs was 251 pairs
more than the original number. The plots of $P_{pair}$, $P_{single}$, and
$P_{none}$ with the corrected parameters are shown in Figure 3. Little
difference was observed in the shape of the $P_{pair}$ curves obtained with
the corrected and uncorrected parameters. The number of lost gene pairs to be
counted was corrected to 1476 (Figure 3B), but the number observable in the
extant genomes was estimated to be 1209 according to the above equation, which
was close to the number in the original data.\newline\newline%
\begin{tabular}
[c]{lll}%
\multicolumn{3}{l}{{\small Table 2. Underestimation of gene pairs in the state
\textquotedblleft pair\textquotedblright\ by parallel gene losses.}}\\\hline
{\small Time point of node (}$t${\small )} & {\small Ratio of underestimation
(}$\pm${\small SD)} & ${\small Q}$\\\hline
${\small -245}$ & ${\small 0.093\pm0.0085}$ & ${\small 0.046}$\\
${\small -164}$ & ${\small 0.0098\pm0.0032}$ & ${\small 0.0098}$\\
${\small -158}$ & ${\small 0.0029\pm0.0018}$ & ${\small 0.0022}$\\
${\small -120}$ & ${\small 0.00031\pm0.00059}$ & ${\small 0.00016}$\\
${\small -105}$ & ${\small 0.0021\pm0.0016}$ & ${\small 0.0022}$\\
${\small -86}$ & ${\small 0.00041\pm0.00067}$ & ${\small 0.00033}$\\
${\small -74}$ & ${\small 0.0017\pm0.0014}$ & ${\small 0.0018}$\\
${\small -41}$ & ${\small 0.00051\pm0.00078}$ & ${\small 0.00053}$\\
\multicolumn{3}{l}{}\\
{\small Time of TS-WGD} & {\small Ratio of underestimation (}$\pm${\small SD)}
& ${\small R}$\\\hline
${\small -306}$ & ${\small 0.032\pm0.0022}$ & ${\small 0.031}$\\\hline
\multicolumn{3}{l}{{\footnotesize Simulations were performed 1000 times.}}\\
\multicolumn{3}{l}{{\footnotesize SD, standard deviation.}}\\
\multicolumn{3}{l}{{\footnotesize TS-WGD, teleost-specific whole genome
duplication.}}%
\end{tabular}
\linebreak

In this equilibrium model, the value of $P_{none}$ converges to 1 at
$t=\infty$ ($\gamma>0$), indicating that, theoretically, all the genes will be
lost in the long-term future. Such a prediction seems to be unnatural from the
view of genome evolution. However, it should be noted that the start number of
gene pairs was fixed in the modeling ($N$ = 7143, or originally 6892), and the
gene gain event was not taken into account. It is possible that the number of
genes gained after WGD may compensate for the number lost. In addition, the
value of $\gamma$ is very small, therefore it will take about 700 million
years from the present for the number of genes to decrease by half ($N/2$ =
3571.5) according to the model. This is a long enough time for the gene
content to be influenced by many other evolutionary mechanisms; thus,
ultra-long-term predictions (about 1000 million years after WGD) by the
equilibrium model are not practical. Rather, the equilibrium model estimates
the evolutionary features of duplicate genes in the present or before. Here, I
focused on $P_{d}(t)$, the probability that a pair of genes derived from WGD
differentiated functionally at time $t$ (Figure 1). The ratio of $P_{d}$ to
the probability that a gene pair remains at the present, that is
$P_{d}(306)/P_{pair}(306)$, was almost 1, implying that almost all the gene
pairs present in the extant teleosts have already differentiated functionally.
This conjecture is consistent with the result from a recent gene expression
analysis study \cite{ref17}, but that is based on a limited number of gene
pairs examined only in zebrafish. Further researches using many genes and/or
many teleost genomes will need to be carried out to test the equilibrium model.

The advantage of equilibrium model is that the impact of WGD can be
quantitatively assessed using the model parameters. For example, the value of
$\beta$, which is the rate of functional differentiation in a gene pair, may
be correlated with the occurrence of novel genotypes triggered by WGD. Many
detailed models about the mechanisms of functional differentiation have been
proposed (reviewed in \cite{ref18}), and $\beta$ may be regarded as an
averaged index of the effects of these mechanisms at the genomic level. In
addition, $\beta/\left(  \alpha+\beta\right)  $ indicates the ideal proportion
of functionally differentiated gene pairs out of all gene pairs when $t\gg0$
and $\gamma=0$. Practically, $P_{d}$ ($\gamma>0$) better reflects the
proportion,%
\[
P_{d}(t)=\frac{\beta\left(  e^{-\gamma t}-e^{-\left(  \alpha+\beta\right)
t}\right)  }{\alpha+\beta-\gamma},
\]
\linebreak and the maximum is obtained using the derivative $dP_{d}/dt=0$. For
the teleost data used in the present study, the maximum of $P_{d}(t+306)$ was
about 0.13 with $t=-214$ (Figure 4).%
\begin{figure}
[ptb]
\begin{center}
\includegraphics[
height=2.1828in,
width=2.1655in
]%
{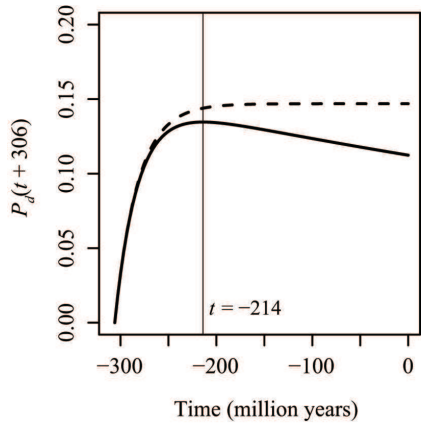}%
\caption{{\protect\small Estimated proportion of gene pairs that have
differentiated functionally in teleost genomes. The }$P_{d}(t+306)$%
{\protect\small \ curves are shown as a solid line (}$\gamma>0$%
{\protect\small ) and a dashed line (}$\gamma=0${\protect\small ). The value
of }$P_{d}(t+306)${\protect\small \ reaches a maximum at }$t=-214$%
{\protect\small \ million years when }$\gamma=0.00095${\protect\small , or
converges to }$\beta/\left(  \alpha+\beta\right)  (\simeq0.15)$%
{\protect\small \ when }$\gamma=0${\protect\small .}}%
\end{center}
\end{figure}
Thus, assuming that the total gene number for standard teleost species is
20000--25000 \cite{ref19,ref20}, a maximum of 2600--3300 gene pairs were
estimated to have differentiated functionally during 90 million years after
TS-WGD. In the case of plant species, it was reported that 99\% of about 2000
duplicate gene pairs that were examined in the cotton genome had
differentiated at the gene expression level during 60 million years after WGD,
and probably those have evolved under purifying selection \cite{ref21}.
Therefore, the estimate of functionally differentiated gene pairs in teleost
species might not be very surprising. It should be stressed that if the
equilibrium model is valid not only for teleosts but also for other lineages
that have undergone WGDs, the above-mentioned indices could be compared among
such lineages. From a na\"{\i}ve perspective, the value of $\beta$ or
$\beta/\left(  \alpha+\beta\right)  $ (or $P_{d}$) may be directly or
indirectly associated with evolutionarily significant events in the lineages
examined, such as the occurrence of novel phenotypes, or other evolutionary
features such as the diversity of species or population size. Regarding the
teleosts, it is known that one more WGD occurred recently (%
$<$%
100 million years ago) in the lineage of Salmonids after the TS-WGD (reviewed
in \cite{ref22}). Therefore, further comparisons using Salmonid genomic data
may allow the impact of WGD to be assessed and compared within the teleost lineage.

\end{document}